\documentclass[10pt,conference,a4paper]{IEEEtran}
\usepackage{times,amsmath,amssymb,epsfig,cite}
\usepackage[latin1]{inputenc}
\newcommand{\diff}[0]{\text{d}}
\newcommand{\im}[0]{\text{Im}\,}
\newcommand{\re}[0]{\text{Re}\,}
\title{Causality and Its Implications for Passive and Active Media}
\author{%
{Johannes Skaar}%
\vspace{1.6mm}\\
\fontsize{10}{10}\selectfont\itshape
Department of Electronics and Telecommunications, Norwegian University of Science and Technology (NTNU),\\ NO-7491 Trondheim, Norway; and University Graduate Center (UNIK), NO-2027 Kjeller, Norway.\\
\\
\fontsize{9}{9}\selectfont\ttfamily\upshape
johannes.skaar@iet.ntnu.no\\
\vspace{1.2mm}\\
\fontsize{10}{10}\selectfont\rmfamily\itshape
}
\begin{document}
\maketitle
\begin{abstract}
The implications of causality are examined in detail for the permittivity, permeability, and refractive index of passive and active media. Particular attention is given to negative index media.
\end{abstract}

%
\section{Introduction}
The ability of fabricating artificial materials with sophisticated electromagnetic properties have generated large interest recently. By tailoring the electric permittivity $\epsilon$ and the magnetic permeability $\mu$, one can obtain arbitrary metamaterial-induced coordinate transformations, leading to devices such as perfect lenses and electromagnetic cloaks \cite{pendry2000,pendry2006,leonhardt2006}.

At a single, positive frequency, a linear medium can have any $\epsilon$ and $\mu$. If the imaginary parts of one or both of these parameters are negative, the medium is active; otherwise it may be passive. When considering global frequency-dependent parameters $\epsilon(\omega)$ and $\mu(\omega)$, there are certain implications and limitations due to causality. We will analyze these connections in more detail, and summarize several results about passive and active media, beyond the usual Kramers--Kronig relations. We emphasize that only causality (and passitivity in the case of a passive medium) are taken into account. Thus the results are fundamental, and not related to practical fabrication problems.

We restrict the attention to linear, isotropic, homogeneous, time-shift-invariant media without spatial dispersion.

\section{Passive media}
The permittivity of a passive medium satisfies the Kramers--Kronig relations \cite{landau_lifshitz_edcm}\footnote{If $\epsilon(\omega)$ has singularities at the real axis, the Kramers--Kronig relations must be modified. For example, if the medium is conducting at zero frequency, $\epsilon(\omega)$ is singular at $\omega=0$. Then, the Kramers--Kronig relations are retained if we subtract the singularity, i.e., make the substitution $\epsilon(\omega)\to\epsilon(\omega)-i\sigma/\omega$ in \eqref{KK}, where $\sigma>0$ is the zero frequency conductivity \cite{landau_lifshitz_edcm}. These media have larger loss for the same $\re\epsilon(\omega)$.}:
\begin{subequations}
\label{KK}
\begin{align}
&\im \epsilon(\omega)=\frac{2\omega}{\pi}\mathcal P\int_{0}^{\infty}\frac{\re\epsilon(\omega')-1}{\omega^2-\omega'^2}d\omega',\label{KK1}\\
&\re \epsilon(\omega)-1=\frac{2}{\pi}\mathcal P\int_0^\infty\frac{\im\epsilon(\omega')\omega'}{\omega'^2-\omega^2}d\omega',\label{KK2}
\end{align}
\end{subequations}
where $\mathcal P$ denotes the Cauchy principal value. Moreover, the permittivity is defined for negative frequencies by the symmetry relation
\begin{equation}
\epsilon(-\omega)=\epsilon^*(\omega), \label{sym}
\end{equation}
so that their inverse Fourier transforms are real. In addition to (\ref{KK})-(\ref{sym}) we have: 
\begin{equation}
\im \epsilon(\omega)>0 \qquad\rm{for}\ \, \omega>0, \label{loss}
\end{equation}
which is the passitivity condition. The losses, as given by the imaginary parts of the permittivity, can be vanishingly small; however they are always present unless we are considering vacuum \cite{landau_lifshitz_edcm}.

In the following we will assume that the permeability $\mu(\omega)$ also satisfies relations (\ref{KK})-(\ref{sym}), and \eqref{loss} for a passive medium, as is usually assumed in the literature. It is to be noted, however, that this is not always exactly true. Indeed, using a Kramers--Kronig relation analogously to \eqref{KK2} to calculate $\re\mu(0)$ from $\im\mu(\omega)$, we straightforwardly find that $\re\mu(0)>1$ whenever $\im\mu(\omega)>0$. This is not always correct \cite{martin67}, as demonstrated by the existence of passive, diamagnetic media.

Eqs. (\ref{KK})-(\ref{loss}) imply that $\epsilon(\omega)$ is analytic and zero-free in the upper half-plane $\im\omega>0$ \cite{landau_lifshitz_edcm}. Thus the refractive index $n(\omega)=\sqrt{\epsilon(\omega)}\sqrt{\mu(\omega)}$ can always be chosen as an analytic function there. With the additional choice that $n(\omega)\to +1$ as $\omega\to\infty$, $n(\omega)$ is determined uniquely, and it follows that (\ref{KK})-(\ref{loss}) hold for the substitution $\epsilon(\omega)\to n(\omega)$ \cite{nussenzveig,SS06}.

We will now consider the implications of causality and passitivity. We will write the conditions in terms of the refractive index $n(\omega)$, although identical conditions apply to the permittivity and the permeability as well.

\subsection{Negative refraction is possible at a single frequency, with arbitrarily low maximum loss for all $\omega$.}
\begin{figure}
  \centering
  \includegraphics{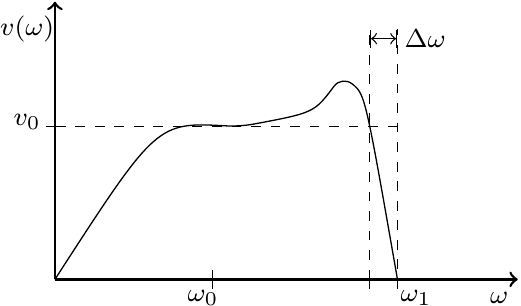}
  \caption{The loss function $v(\omega)$ has support below $\omega_1$. For $\omega_0<\omega<\omega_1-\Delta\omega$, we assume $v(\omega)\geq v_0$.}
  \label{fig:lowbound}
\end{figure}

\noindent
The proof is by construction. Let the refractive index  be written $n(\omega)=1+u(\omega)+iv(\omega)$, where $1+u(\omega)$ and $v(\omega)$ are the real and imaginary parts, respectively. Passitivity means that $v(\omega)>0$ for $\omega>0$. Since $v(\omega)=0$ can be approached, we allow ourselves to put $v(\omega)=0$ as well; adding a small, slowly varying function to $v(\omega)$ does not alter the argument below. Let $v(\omega)\geq v_0>0$ for $\omega_0<\omega<\omega_1-\Delta\omega$, and $v(\omega)=0$ for $\omega>\omega_1$, see Fig. \ref{fig:lowbound}. 
Estimating $u(\omega_1)$ by the Kramers--Kronig relation \eqref{KK2}, we find
\begin{eqnarray}
u(\omega_1)&&=-\frac{2}{\pi}\int_0^{\omega_1}\frac{v(\omega)\omega d\omega}{\omega_1^2-\omega^2}\nonumber\\
&&\leq -\frac{v_0}{\pi}\int_{\omega_0}^{\omega_1-\Delta\omega}\frac{2\omega d\omega}{\omega_1^2-\omega^2} \nonumber\\
&&= -\frac{v_0}{\pi}\left(\ln\frac{\omega_1-\omega_0}{\Delta\omega}
-\ln\frac{2\omega_1-\Delta\omega}{\omega_0+\omega_1}\right).
\label{upperbound}
\end{eqnarray}
Now we can choose any small $v_0$ and require the maximum of $v(\omega)$ to be less than, say $2v_0$. By having a sufficiently narrow transition region $\Delta\omega$, $u(\omega_1)$ gets less than $-1$, which completes the proof.

Note that the required function can be approached by superpositions of several, narrow Lorentzians, with resonance frequencies equally spaced in the interval $(\omega_0,\omega_1-\Delta\omega)$. For example, in the limit of continuous varying resonance frequencies from $\omega_0=0$ to $\omega_1-\Delta\omega$, we obtain
\begin{eqnarray}
\label{lorentzsuperposition}
 u(\omega)+iv(\omega) &&\propto \int_0^{\omega_1-\Delta\omega} \frac{\omega_0^2d\omega_0}{\omega_0^2-\omega^2-i\omega\Gamma}\\
&&=\omega_1-\Delta\omega - s\arctan\left(\frac{\omega_1-\Delta\omega}{s}\right), \nonumber 
\end{eqnarray}
where $s=i\sqrt{\omega^2+i\omega\Gamma)}$. The required refractive index function is obtained by choosing a sufficiently small $\Gamma$.

The steep edge in the transition band may imply that the medium is difficult to realize. Given a small maximum loss, $v(\omega)\leq v_{\max}$ for all $\omega$, one can prove that such steep edges are the only ways to obtain negative refraction for passive media that satisfy Kramers--Kronig relations in the usual form \eqref{KK}. Indeed, for any square integrable function $v(\omega)\geq 0$ with limited steepness ($|dv(\omega)/d\omega|\leq 1/\Delta\omega$ for some $\Delta\omega$),
\begin{equation}
\label{lowerbound}
u(\omega_1)\geq -\frac{v_{\max}}{\pi}\left(\ln\frac{2\omega_1}{v_{\max}\Delta\omega}
+1\right).
\end{equation}
Here $\omega_1$ is an observation frequency. The inequality \eqref{lowerbound} is found by a similar argument as that of \eqref{upperbound}, considering the fact that the least possible $u(\omega_1)$ is obtained when $v(\omega)=v_{\max}$ for $0<\omega\leq\omega_1-\delta\omega$, and $v(\omega)$ decreases linearly to zero above $\omega_1-\delta\omega$. Here $\delta\omega$ is a positive parameter. For media with singularities of $\epsilon(\omega)$ or $\mu(\omega)$ at real frequencies (e.g. an ideal plasma), \eqref{lowerbound} does not apply. We can conclude:
\subsection{Passive, negative refraction implies either (i) large loss below the working frequency, or (ii) exponentially steep variation immediately below the working frequency, or (iii) singularities at real frequencies.} 
\noindent
The trade-off between requirements (i) and (ii) is quantified by \eqref{lowerbound}.

\subsection{Negative refraction is possible in any finite bandwidth, with arbitrarily low loss there.}
\noindent
This fact follows from a similar argument as that of the estimation \eqref{upperbound}; the trick is to choose a loss function that is sufficiently large below the interval of interest, and zero or sufficiently small in and above the interval.

\subsection{If a medium is lossless in a finite bandwidth $[\omega_1,\omega_2]$, the minimum variation of $n$ in this bandwidth is}\noindent
\begin{equation}
 |\re n(\omega_1)-1|(2\Delta-\Delta^2)  \text{ for }  n(\omega_1)<1.
\end{equation}
Here we have defined the normalized bandwidth
\begin{equation}
\Delta=(\omega_2-\omega_1)/\omega_2.
\end{equation}
For $n(\omega_1)\ge 1$ there is no lower bound for the variation of $n(\omega)$.

In particular, the derivative $\diff n/\diff\omega$ is bounded from below:
\begin{equation}
\label{dnbound}
\frac{\diff n}{\diff\omega} >
\begin{cases}
2|n(\omega)-1|/\omega & \text{ for } n(\omega)<1, \\
  0 & \text{ for } n(\omega)\geq 1. 
\end{cases}
\end{equation}

The proof is relatively simple, based on the Kramers--Kronig relations \cite{SS06}.  

\subsection{If $\re n(\omega)$ is constant and less than unity in $[\omega_1,\omega_2]$, there is a frequency in $[\omega_1,\omega_2]$ with loss}\noindent
\begin{align}
\im n(\omega)&>|\re n(\omega_1)-1|\frac{\omega_2^2-\omega_1^2}{2\omega_1\omega_2}\nonumber\\
&=|\re n(\omega_1)-1|\Delta + O(\Delta^2) 
\end{align}
Thus, a constant $\re n(\omega)$ in a finite bandwidth, less than unity, is not possible unless the loss is sufficiently large. The proof is somewhat complicated \cite{SS06,SS05}.

\subsection{Consider a passive, so-called perfect lens \cite{pendry2000}. If the lens is used for frequencies in a finite bandwidth $[\omega_1,\omega_2]$, the spatial resolution is $2\pi d/|\ln(\Delta/2)|$.}
\noindent
Here we have quantified the worst resolution with respect to frequency in the interval, and optimized the medium subject to \eqref{KK}-\eqref{loss}. Thus, for a linear improvement of the resolution, the bandwidth must be shrinked exponentially. The proof is given in \cite{LJSS09}.
 
\section{Active media}
The permittivity of an active medium satisfies \eqref{KK} and \eqref{sym}, but not \eqref{loss}. Similar relations are valid for the permeability. Determining the sign of the refractive index $n(\omega)=\sqrt{\epsilon(\omega)\mu(\omega)}$ is not straightforward, as it is for passive media. For example, for a certain frequency it is no longer true to say that causality dictates a certain sign of $n$, based on $\epsilon$ and $\mu$ at this frequency. One really has to go back to first principles: Causality means that no signal can go faster than the speed of light in vacuum. Restricting ourselves to media with no poles or odd-order zeros of $\epsilon(\omega)\mu(\omega)$ in the upper half-plane, it can be shown that causality implies that $n(\omega)$ must be identified as an analytic function in the upper half-plane, with asymptotic behavior $n(\omega)\to+1$ as $\omega\to\infty$ \cite{skaar06,nistad08}. This is in accordance with the classical result for passive media \cite{brillouin}. 

In practice, the sign of the refractive index is conveniently determined using the following result:
\subsection{If $\epsilon(\omega)\mu(\omega)$ is continuous and zero-free for real frequencies (except possibly at $\omega=0$), and analytic and zero-free in the upper half-plane, the refractive index is given by}
\noindent 
\begin{equation}
n(\omega)=\sqrt{|\epsilon(\omega)||\mu(\omega)|}\exp[i(\varphi_\epsilon(\omega)+\varphi_\mu(\omega))/2)],
\label{eq:ndef}
\end{equation}
where $\varphi_\epsilon(\omega)+\varphi_\mu(\omega)$ is the complex argument of $\epsilon(\omega)\mu(\omega)$, unwrapped such that it is continuous for $\omega\neq 0$ and such that it tends to $0$ as $\omega\to\pm\infty$ \cite{skaar06b}.

The phase unwrapping procedure indicates that the sign of the refractive index is not uniquely determined from $\epsilon$ and $\mu$ at a single frequency; knowledge of the global functions $\epsilon(\omega)$ and $\mu(\omega)$ are required to determine the sign, even at a single frequency:
\subsection{Two media with identical $\epsilon$ and identical $\mu$ at a certain frequency may have different $n$ at this frequency.}
\noindent
Consider two nonmagnetic materials with $\epsilon_1(\omega)=1+\chi_g(\omega)$ and $\epsilon_2(\omega)=[1+\chi_r(\omega)]^2$, respectively, where $\chi_{r,g}(\omega)$ are Lorentzians in the form 
\begin{equation}\label{lorentzian}
\chi_{{r,g}}(\omega)=\frac{F_{r,g}\omega_{r,g}^2}{\omega_{r,g}^2-\omega^2-i\omega\Gamma_{r,g}}.
\end{equation}
With the parameters $F_g=-0.00016$, $\Gamma_g=0.005$, and $\omega_g=1$ for material 1, and $F_r=3.7$, $\Gamma_r=0.005$, and $\omega_r=0.6$ for material 2, these two materials have $\epsilon_{1,2}(\omega_g)=1-0.030i$. With the help of \eqref{eq:ndef}, we find the refractive indices $n_1(\omega_g)\approx1-0.015i$ and $n_2(\omega_g)\approx-1+0.015i$. 

As $\mu=1$ in the above examples, we obtain the following result:
\subsection{There exist right-handed negative index media.}
\begin{figure}[t!]
\includegraphics[height=5.5cm,width=7cm]{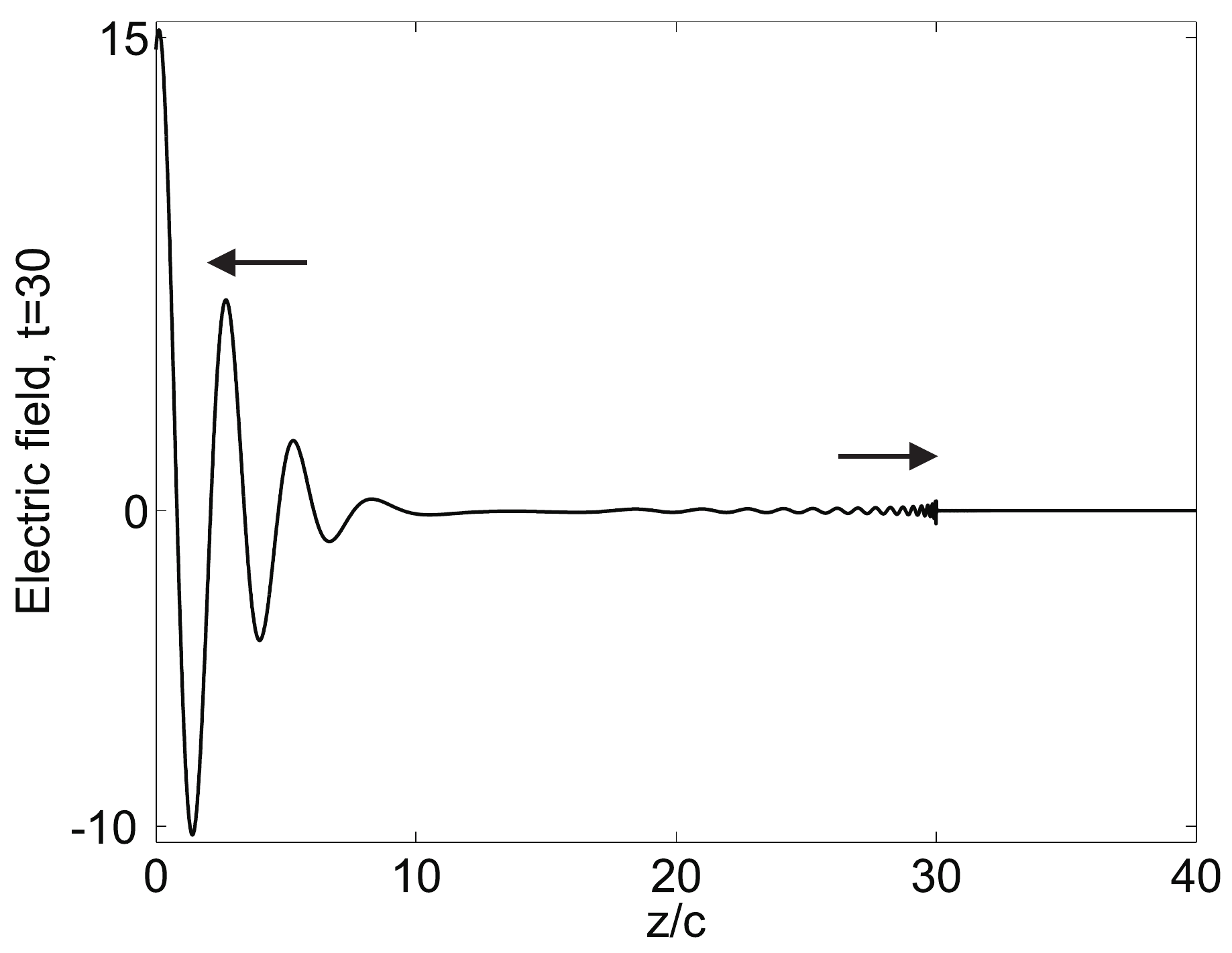}
\caption{The electric field in a right-handed negative index medium. The medium fills the half-space $z\geq 0$, while there is vacuum for $z<0$. A unit-step modulated cosine is incident from vacuum, and the plot is a snapshot at time $t=30$. For details see Ref. \cite{skaar06}. Animations are given in Ref. \cite{nistad07}.}
\label{fig:backwave}
\end{figure}

\noindent 
Also, there exist left-handed positive index media. These facts were first demonstrated in Refs. \cite{chen05,chen06}. 
It is useful to examine the right-handed negative index media in more detail. Let a unit-step modulated cosine wave be incident to a half-spaced filled with such a medium. The time-domain field after some time is given in Fig. \ref{fig:backwave}. The field plot for small $z$ shows the ``backward wave'', with phase velocity and Poynting's vector in the $-z$-direction. This wave draws energy from the medium and grows in the $-z$-direction. As always, the precursor, located close to $z=ct$, moves at a speed $c$ in the $+z$-direction.

\subsection{Define the squared longitudinal wavenumber $k_z^2=\epsilon(\omega)\mu(\omega)\omega^2-k_x^2-k_y^2$. If $\epsilon(\omega)\mu(\omega)\omega^2-k_x^2-k_y^2$ has poles or odd-order zeros in the upper half-plane, $k_z$ does not have physical meaning for real frequencies.}
\noindent
As a special case, if $\epsilon(\omega)\mu(\omega)$ has poles or odd-order zeros in the upper half-plane, the refractive index $n(\omega)$ does not have physical meaning for real frequencies. Although such media may be causal, they do not obey the conventional Kramers--Kronig relations for real frequencies; we must let the frequencies be complex with imaginary values larger than those of all nonanalytic points of $n$.

The above result follows from the fact that the physical fields are the real time-domain fields. To obtain the time-domain fields from the frequency-domain fields, one performs an inverse Fourier or Laplace transform, that is, an integral above all nonanalytic points in the upper half-plane. Performing the integral along the real axis leads necessarily to unphysical results as the integral becomes dependent on the choice of branch cut locations \cite{skaar06}.

When constructing active metamaterials, one should examine whether $\epsilon(\omega)\mu(\omega)$ has poles or odd-order zeros in the upper half-plane. If there are such zeros or poles, the medium has so-called absolute unstabilities, which makes it useless for linear, small-signal applications. An absolute instability means that the fields blow up at any fixed point in space; the pulse is not convected away as in conventional gain media \cite{sturrock,briggs,skaar06,nistad08}. 

It is to be noted that for oblique incidence, even conventional, weak gain media (such as Erbium-doped glasses) may have nonanalytic points of $k_z$ in the upper half-plane \cite{skaar06b}. Thus for oblique incidence one must be particularly careful with the use or interpretation of $k_z$.


\subsection{A causal refractive index function $n(\omega)$ can approximate any square integrable function $f(\omega)$ on a finite bandwidth.}
\noindent
The approximation may be achieved with any precision\footnote{In mathematical terms, a function $h\in H^2$ satisfying $h(-\omega^*)=h^*(\omega)$ can approximate any function $f\in L^2(\omega_1,\omega_2)$, as precisely as desired in the corresponding metric. Here $H^2$ denotes the Hardy space of the upper half-plane, and $0\leq\omega_1<\omega_2<\infty$. See e.g. Ref. \cite{KrNu2}.}; in the Kre{\u\i}n--Nudel$'$man case \cite{KrNu2} at the expense of the norm of $n(\omega)-1$ outside the bandwidth of interest. Of course, a large norm outside the relevant bandwidth may imply difficulties of realization. Nevertheless, we note that causality does not prohibit e.g. $n(\omega)\approx -1$ with any precision, even in a finite bandwidth. The possibility of approximating any desired behavior in a limited bandwidth may seem useless unless the resulting medium is free from absolute instabilities. Fortunately, there is a remedy in that $\log f(\omega)$ can be approximated by a function $g(\omega)$ using Kre{\u\i}n--Nudel$'$man, and setting $\epsilon(\omega)=\exp[g(\omega)]$ \cite{skaar01}.
 
As an example of gain compensation of the losses associated with a left-handed resonance, consider the causal medium $\epsilon(\omega)=1+\chi_r(\omega)+\chi_g(\omega)$ and $\mu(\omega)=1+\chi_r(\omega)$, where $\chi_{r,g}(\omega)$ are the Lorentzians defined in \eqref{lorentzian}. Taking $\omega_r=1$, $\omega_g=1.058$, $F_r=0.25$, $F_g=-0.0034$, $\Gamma_r=0.005$, and $\Gamma_g=0.02$, we find that $n(\omega)=-1$ and $\diff\im n^2(\omega)/\diff\omega=0$ for $\omega=1.06$.
\begin{figure}
  \centering
  \includegraphics{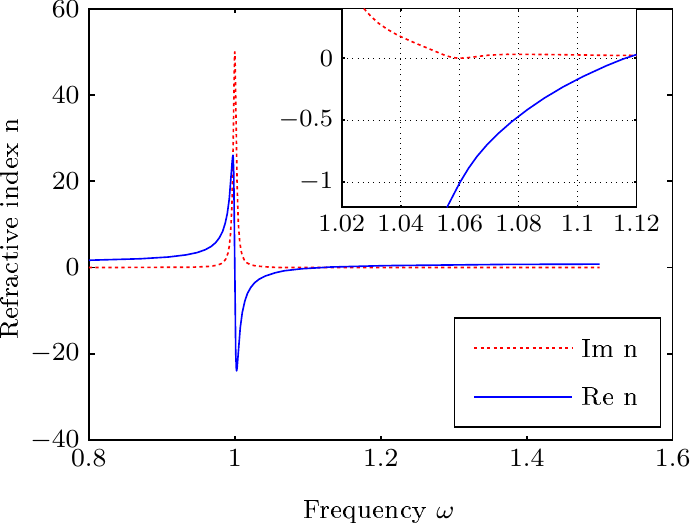}
  \caption{Refractive index for the left-handed medium with gain compensation. At $\omega=1.06$ we have lossless negative refraction, $n=-1$ and $\diff\im n^2(\omega)/\diff\omega=0$.}
  \label{fig:stockman}
\end{figure}
This medium has $\im\epsilon(\omega)<0$ in the bandwidth $[1.05,1.07]$; thus it is net active there. This does not imply that the system is unstable unless the medium is infinite or put in a resonator configuration; examples of such stable systems include fiber optic amplifiers. The absence of absolute instabilities of this medium is guaranteed by the fact that $\epsilon(\omega)$ and $\mu(\omega)$ do not contain zeros or poles in the upper half-plane. We note that the gain compensation due to $\chi_g(\omega)$ has completely removed the loss which would have been present in the absence of this gain.
 

\section{Conclusion}
For passive media, we have first stated that causality does not prohibit negative refraction with arbitrarily low losses, even in a finite bandwidth. Negative refraction implies either (i) large loss below the working frequency, or (ii) exponentially steep variation immediately below the working frequency, or (iii) singularities at real frequencies. However, for a fixed $\omega$ there is a lower bound for the loss $\im n(\omega)$ if $\diff\re n(\omega)/\diff\omega=0$, or lower bound for $|\diff\re n(\omega)/\diff\omega|$ if $\im n(\omega)=0$. 

For active media, there are no fundamental limitations in a finite bandwidth. Nevertheless, there are some peculiar results related to the sign of the refractive index, and backward waves. Two media with identical $\epsilon$ and $\mu$ at a certain frequency $\omega$ can have different $n$. The sign of $n(\omega)$ must be determined from the global frequency-dependence of $\epsilon(\omega)$ and $\mu(\omega)$. If active media are to be used in linear applications, it is cruical to eliminate the presence of absolute instabilities; otherwise the refractive index does not have physical meaning for real frequencies.

\section*{Acknowledgment}
Kristian Seip and Bertil Nistad are acknowledged for helpful discussions.

\bibliographystyle{IEEEtran}
\bibliography{nbib}

\end{document}